\documentclass[12pt]{article}

\begin{document}

\title{Does The Force From an Extra Dimension Contradict Physics in $4D$?}
\author{J. Ponce de Leon\thanks{E-mail: jponce@upracd.upr.clu.edu or jpdel@astro.uwaterloo.ca}\\ Laboratory of Theoretical Physics, Department of Physics\\ 
University of Puerto Rico, P.O. Box 23343, San Juan, \\ PR 00931, USA \\ 
Department of Physics, University of Waterloo,\\
 Waterloo, Ontario N2L 3G1, Canada.}
\date{October 2001}

\maketitle
\begin{abstract}
We examine the question of whether violation of $4D$ physics is an inevitable consequence of existence of an extra non-compactified dimension. Recent investigations in membrane and Kaluza-Klein theory indicate that when the metric of the spacetime is allowed to depend on the extra coordinate, i.e., the cilindricity condition is dropped, the equation describing the trajectory of a particle in one lower dimension has an extra force with some abnormal properties. Among them, a force term parallel to the four-velocity of the particle and, what is perhaps more surprising, $u_{\mu}f^{\mu} \neq u^{\mu}f_{\mu}$. These properties violate basic concepts in $4D$ physics. In this paper we argue that these abnormal properties are {\em not} consequence of the extra dimension, but result from the formalism used. We propose a new definition for the force, from the extra dimension, which is free of any contradictions and consistent with usual $4D$ physics. We show, using warp metrics, that this new definition is also more consistent with our physical intuition. The effects of this force could be detected observing objects moving with high speed, near black holes and/or in cosmological situations. 

 \end{abstract}



\newpage

\section{INTRODUCTION}
In ``classical" versions of Kaluza-Klein theory the so-called cylinder condition is one of the basic assumptions. This condition basically states that metric coefficients do not depend on the fifth coordinate, in such a way that all derivatives with respect to this coordinate vanish. 

In the last years there is a consensus in the physics community that cilindricity condition is not required nor, in general, sustained. Indeed, it is now a common assumption that the metric tensor as well as other physical quantities depend on the fifth coordinate \cite{Wesson book}-\cite{Deruelle and Katz}. In multidimensional theories called ``brane-world" models as well as in the space-time-matter theory in $5D$, the implications of such an assumption are presently under intensive theoretical study.

In particular, the effects of extra dimensions on the trajectory of tests particles, as observed in $4D$ (one lower dimension) have been studied \cite{Youm}, \cite{Wesson Mashhonn}. Employing  techniques similar to the ones used in classical Kaluza-Klein theory, a number of results have been obtained. For example, the dependence of the metric on the extra coordinate leads to, a new force term which presents two important properties, viz., (i) it is proportional to the first derivative of the metric with respect to the extra coordinate and (ii) it has a component  which is parallel to the four-velocity of the particle. 

The fist property implies that the extra force cannot be implemented directly in brane-world models, in the RS2 scenario \cite{RS2}. This is because these derivatives are discontinuous, and change sign, through the brane due to the $\delta$-function singularity there. However, as it is discussed in \cite{Youm 2}, effective $4D$ equations can be obtained by taking mean values and applying Israel's junction conditions through the brane. The resulting effective extra force depends on whether the brane universe is invariant or not under the ${\bf Z}_2$ transformation. In a more realistic theory where the brane is not assumed to be infinitely thin, but has a finite width determined by the specifics of the theory, the extra force should be continuous, changing its sign, as one moves through the brane. Thus, for such ``thick" branes there should be a region, near the core of the brane where the force vanishes identically.

In this paper we deal with the second property mentioned above. Namely, that the extra force has a component parallel to the four-velocity of the particle. This is a violation of the laws of physics in $4D$, where the 4-velocity $u^{\mu}$ and the 4-force are always orthogonal. Even more astonishing is the fact that $u_{\mu}f^{\mu} \neq u^{\mu}f_{\mu}$, which makes even harder the physical interpretation of this force. Due to these unusual properties, which cannot be explained by conventional $4D$ physics, such extra force has been called fifth force \cite{Wesson book}. 

Does the force, from an extra dimension, necessarily violate physics in $4D$? The current answer in the literature is positive \cite{Youm}, \cite{Wesson Mashhonn}, \cite{Youm 2}. This is an important question, from a theoretical and observational/experimental point of view. Therefore, it should be thoroughly investigated, from different angles and perspectives.

The aim of this paper is to provide a less radical answer to this question. Namely, that the force from an extra dimension {\em does not} necessarily contradict $4D$ physics. Our interpretation is that the abnormal properties of the fifth force are consequence of the formalism used. 

First, we will see that when the metric is allowed to depend on the extra coordinate, the formalism and definitions used in classical Kaluza-Klein theory are incompatible with the requirement of gauge invariance. 

Second, we will show how to introduce  a new definition for the $4D$ force, from an extra dimension, which is free of any contradictions and consistent with usual $4D$ physics.

\section{Line Element in Kaluza-Klein Theory}
To facilitate the discussion and set the notation, we start with a brief summary of the Kaluza-Klein equations. We consider a five-dimensional manifold with coordinates $\xi^{A}$ $(A = 0,1,2,3,4)$ and metric tensor $\gamma_{AB}$. The $5D$ interval is then given by
\begin{equation}
\label{general 5D metric}
d{\cal S}^2 = \gamma_{AB}d{\xi}^A d{\xi}^B.
\end{equation}
It is a popular choice to consider that the first four coordinates ${\xi}^\mu$ are the coordinates of the spacetime $x^{\mu}$ $(\mu = 0,1,2,3)$, while ${\xi}^4$ is the extra dimension, which we will denote $y$, viz,
\begin{eqnarray}
x^\mu = {\xi}^{\mu}\nonumber \\
y = \xi^4.
\end{eqnarray}
Now setting $\gamma_{\mu4} = \gamma_{44}A_{\mu}$ and $\gamma_{44} = \epsilon\Phi^2$, the general line element (\ref{general 5D metric}), without any loss of generality, can be written as
\begin{equation}
\label{5D metric in special coordinates}
d{\bf {\cal S}}^{2} = ds^{2} + \epsilon \Phi^{2}(dy + A_{\mu}
dx^{\mu})^{2},
\end{equation}
 where $ds^{2} = g_{\mu \nu} dx^{\mu} dx^{\nu}$ is the spacetime interval with metric $g_{\mu\nu} = (\gamma_{\mu\nu} - \epsilon \Phi^2 A_{\mu}A_{\nu})$. The quantities $\Phi$ and $A_{\mu}$ are called the scalar and vector potentials, respectively. The factor $\epsilon$ is taken to be $+1$ or $-1$ depending on whether the extra dimension is timelike or spacelike, respectively.
The above separation is invariant under the set of transformations
\begin{eqnarray}
\label{Allowed transformations}
x^{\mu} &=& \bar{x}^{\mu},\nonumber \\ y&=& \bar{y}+ f(\bar{x}^{0},\bar{x}^{1},\bar{x}^{2},\bar{x}^{3}),
\end{eqnarray}
which in $5D$ reflect the freedom in the choice of origin for $y$, while in $4D$ correspond to the usual gauge freedom of the potentials 
\begin{equation}
\label{gauge transformations}
\bar{A}_{\mu} = A_{\mu} + \frac{\partial{f}}{\partial{\bar{x}^{\mu}}}= A_{\mu}+ f,_{\mu}.
\end{equation}.

The basic postulate, regarding the question discussed here, is that the equations of motion for test particles are obtained by minimizing interval (\ref{general 5D metric}), or (\ref{5D metric in special coordinates}) in more familiar notation. This postulate, which means that the motion of test particles is geodesic, as well as equations (\ref{general 5D metric})-(\ref{gauge transformations}), are accepted in both, compactified and non-compactified Kaluza-Klein theories. 
\section{Test Particles in Kaluza-Klein Theory}
In this section we critically review the notions that lead to a fifth force. We compare the formalism in the compactified and non-compactified versions of the theory. 
We show that when the definition of force, used in the compactified version, is extended to the non-compactified version we obtain a force which is not gauge invariant. We then discuss the properties of the fifth force.

\subsection{Compactified extra Dimension}
This is the classical Kaluza-Klein theory where physical quantities are allowed to depend on $x^\mu$ but not on $y$ (cylinder condition). The geodesic equation splits up in two sets of equations. The first one, corresponds to the motion in spacetime, and provides a definition for the ``extra" force (per unit mass), namely,   
\begin{equation}
\label{def of force in classical KK}
\frac{Du^\mu}{ds} = \frac{d^2 x^{\mu}}{ds^2} + {\Gamma}^{\mu}_{\alpha \beta}\frac{dx^{\alpha}}{ds}\frac{dx^{\beta}}{ds} = f^{\mu},
\end{equation}
where 
\begin{equation}
\label{explicit form of force in classical KK}
f^{\mu} = (\Phi u^{(4)}){F}^{\mu\rho}u_{\rho} +  
\frac{\epsilon (u^{(4)})^2}{\Phi}\left[\Phi^{\mu} - u^{\mu}\Phi_{\rho}u^{\rho} \right],
\end{equation}
${\Gamma}^{\mu}_{\alpha \beta}$ are the usual Christoffel symbols constructed from $g_{\mu\nu}$, $F_{\mu\nu}$ is the antisymmetric tensor $F_{\mu\nu} = A_{\nu,\mu} - A_{\mu,\nu}$ and $u^{(4)}$ is 
\begin{equation}
u^{(4)}= \epsilon \Phi[\frac{dy}{ds} + A_{\mu}u^{\mu}].   
\end{equation}
The evolution of this quantity is provided by the remaining component of the geodesic equation. It is 
\begin{equation}
\label{equation for u4 in the coordinate frame}
\frac{1}{\left[{{1 + \epsilon (u^{(4)})^2}}\right]}\frac{du^{(4)}}{ds}= - \frac{\Phi_{\mu}}{\Phi}u^{\mu}u^{(4)}.
\end{equation}
All these equations are invariant under the set of gauge transformations (\ref{Allowed transformations}). In particular the force (\ref{def of force in classical KK}), (\ref{explicit form of force in classical KK}) is gauge invariant and orthogonal to the four-velocity $u^{\mu}$, i.e., 
\begin{eqnarray}
\label{self-consistency}
u_{\mu}f^{\mu} = u^{\nu}f_{\nu} = 0\nonumber \\ 
f_{\nu} = g_{\nu\mu}f^{\mu}\nonumber \\
Dg_{\mu\nu} = 0
\end{eqnarray} 

\subsection{Non-compactified Extra Dimension}

This is the typical scenario in membrane theory and Kaluza-Klein gravity. Here the spacetime metric and the other quantities are allowed to be functions of $y$. As before, the $5D$ geodesic equation separates into a $4D$ part and an extra part. 

Again the $4D$  part (\ref{def of force in classical KK}) is used to define the extra force. However, now this definition is not gauge invariant. This is a consequence of the non-invariance of Christoffel symbols under transformations  (\ref{Allowed transformations}), viz.,
\begin{equation}
\label{transformation of Christoffel symbols}
 \bar{\Gamma}^{\lambda}_{\alpha\beta}= {\Gamma}^{\lambda}_{\alpha\beta} + \frac{1}{2}g^{\lambda\rho}(g_{\rho\alpha,y}f_{,\beta} + g_{\rho\beta,y}f_{,\alpha} - g_{\alpha\beta,y}f_{,\rho}), 
\end{equation}
which follows from the fact that $\bar{g}_{\mu\nu,\lambda} = g_{\mu\nu,\lambda} + g_{\mu\nu,y}f_{,\lambda}$. 

Our first conclusion, therefore, is that the definition for the force (\ref{def of force in classical KK}) {\em is inappropriate}, for the general $5D$ metric (\ref{5D metric in special coordinates}). Indeed, a more detailed analysis indicates that, invariance of $4D$ physics under transformations in $5D$ requires changing the usual definition of various quantities, including Christoffel symbols and the electromagnetic tensor $F_{\mu\nu}$. The appropriate definitions are provided in Ref. \cite{Ponce de Leon 1}.

Inspection of (\ref{transformation of Christoffel symbols}) shows that the non-invariance of Christoffel symbols is a result of the inclusion of electromagnetic potentials $A_{\mu}$. These symbols would be gauge invariant if $A_{\mu}$ were zero. This leads to the question of whether the force definition (\ref{def of force in classical KK}) would still work for the simplified metric
\begin{equation}
\label{5D metric restricted metric}
d{\bf {\cal S}}^{2} = g_{\mu \nu}(x^\rho,y) dx^{\mu} dx^{\nu} + \epsilon \Phi^2(x^\rho,y) dy^{2},
\end{equation}
In this case we obtain
\begin{equation}
\label{4D part in noncompact KK}
\frac{Du^{\sigma}}{ds} =  
\epsilon \Phi (\frac{dy}{ds})^2\left[\Phi^{\sigma} - u^{\sigma}\Phi_{\rho}u^{\rho} \right] + \left(\frac{1}{2}u^{\sigma}u^{\lambda} - g^{\sigma\lambda}\right)u^{\rho} \frac{\partial {g_{\lambda \rho}}}{\partial y}\frac{dy}{ds}.
\end{equation}
The first term, representing the force associated with the scalar field $\Phi$, is identical to the one in (\ref{explicit form of force in classical KK}) and satisfies all the appropriate requirements. Therefore, in what follows we will set $\Phi = 1$ and concentrate our attention in the other terms.

The second term in (\ref{4D part in noncompact KK}) behaves like a $4D$ vector under transformations $x^{\mu} = \bar{x}^{\mu}(x^\lambda)$, $y= \bar{y}$ which leave the separation (\ref{5D metric restricted metric}) invariant. This vectorial behavior, apart from (\ref{def of force in classical KK}), is probably the motivation to identify this term with the force (per  unit mass) associated with the existence of a non-compactified extra dimension, viz., 
\begin{equation}
\label{f(lit) contr}
\frac{Du^{\mu}}{ds} = f^{\mu}_{(lit)} = \left(\frac{1}{2}u^{\mu}u^{\lambda} - g^{\mu\lambda}\right)u^{\rho} \frac{\partial {g_{\lambda \rho}}}{\partial y}\frac{dy}{ds}.
\end{equation}
Here $f_{(lit)}^{\mu}$ stands for: force as defined in the literature. This is the so-called fifth force (per unit inertial mass) typical of membrane theory and Kaluza-Klein theory \cite{Youm}, \cite{Wesson Mashhonn}.
\subsubsection{Properties of The Fifth Force}

In order to isolate some of the properties of $f_{(lit)}^{\mu}$, we evaluate $Du_{\mu}/ds$ in an independent way. Omitting intermediate calculations, we obtain

\begin{equation}
\label{f lit cov}
\frac{Du_{\sigma}}{ds} = f_{(lit)\sigma} =  \frac{1}{2}u_{\sigma}u^{\lambda}u^{\rho}\frac{\partial {g_{\lambda\rho}}}{\partial y}\frac{dy}{ds},
\end{equation}
where, for the same reasons as above, we have made the identification with the covariant component of the force. 

The unique properties of $f^{\mu}_{(lit)}$ are immediately obvious. First, it not only has a component parallel to the $4$-velocity of the particle, but also $u_{\mu}f^{\mu}_{(lit)} \neq u^{\nu}f_{(lit)\nu}$, namely,     

\begin{equation}
\label{escalar 1}
u_{\sigma}f^{\sigma}_{(lit)} = - \frac{1}{2}u^{\rho}u^{\lambda}\frac{\partial {g_{\lambda\rho}}}{\partial y}\frac{dy}{ds},
\end{equation}
  
\begin{equation}
\label{escalar 2}
u^{\sigma}f_{(lit)\sigma} = + \frac{1}{2}u^{\rho}u^{\lambda}\frac{\partial {g_{\lambda\rho}}}{\partial y}\frac{dy}{ds}.
\end{equation}
Also,
\begin{equation}
\label{relation between covariant and contravariant components of force in literature}
f_{(lit)\mu}= g_{\mu\sigma}f_{(lit)}^{\sigma} + u^{\rho} \frac{\partial{g_{\mu\rho}}}{\partial y}\frac{dy}{ds}.
\end{equation}
These expressions indicate that, although $f^{\mu}_{(lit)}$ transforms like a four-vector, it is not a ``regular" $4$-vector. Indeed, the above relations are inconsistent with what we usually understand as a $4$-vector. In addition, (\ref{escalar 1}) and (\ref{escalar 2}) seem to contradict each other. Only 
when there is no dependence on $y$ we recover total self-consistency as in (\ref{self-consistency}).

The current interpretation is that the abnormal properties of this force, which violate the laws of $4D$ physics, are consequence of the existence of extra non-compactified dimensions \cite{Youm}, \cite{Wesson Mashhonn}. It is therefore suggested that the Kaluza-Klein scenario can be tested by detecting {\em inconsistencies} with $4D$ physics.  

\section{New Approach. No Contradictions With $4D$ Physics}

In this Section we propose an alternative, less radical, point of view. Our proposal consists of two parts. 

The first part, is that when the condition of cilindricity is dropped, (\ref{def of force in classical KK}) does not constitute a consistent definition of force neither for the general metric (\ref{5D metric in special coordinates}) nor for the simplified one (\ref{5D metric restricted metric}). Only in the classical Kaluza-Klein theory, with cilindricity,  (\ref{def of force in classical KK}) provides a consistent definition of force per unit mass. 
Therefore, the abnormal properties of the extra force discussed above are not a consequence of the extra dimension, but a result of an incorrect definition of force in $4D$.

 The second part is a constructive one. We show how to introduce a new definition for the $4D$ force, which is mathematically correct,  and leads to an extra force, from the extra dimension, which is free of any contradictions and consistent with usual $4D$ physics.

It is not difficult to see that the source of  inconsistencies (from $4D$ viewpoint) in (\ref{escalar 1})-(\ref{relation between covariant and contravariant components of force in literature}) is that now $Dg_{\mu\nu} \neq 0$, instead of $Dg_{\mu\nu} = 0$ as in (\ref{self-consistency}).
\begin{equation}
\label{Abs. Dif. of metric in 5D}
\label{D4 of metric tensor}
Dg_{\mu\nu}= \left[g_{\mu\nu,\rho} - \left({\Gamma}^{\lambda}_{\mu\rho}g_{\lambda\nu}+ {\Gamma}^{\lambda}_{\nu\rho}g_{\lambda\mu}\right) \right]dx^{\rho} + \frac{\partial{g_{\mu\nu}}}{\partial y}dy.
\end{equation}
The first term is the absolute differential in $4D$, which we will denote as $D^{(4)}$. For which $D^{(4)}g_{\mu\nu} = 0$. For an arbitrary vector $V_{\alpha}$
\begin{equation}
\label{separation of D}
DV_{\alpha}= \left(V_{\alpha,\rho} - {\Gamma}^{\lambda}_{\alpha\rho}V_{\lambda}\right)dx^{\rho} + \frac{\partial{V_{\alpha}}}{\partial y}dy = D^{(4)}V_{\alpha} + \frac{\partial{V_{\alpha}}}{\partial y}dy,
\end{equation}
where $D^{(4)}V_{\alpha}$ represents the absolute differential of $V_{\alpha}$ in $4D$.
Obviously, for any object we can define its four-dimensional absolute derivative as 
\begin{equation}
\label{def of D in 4D}
D^{(4)}(\cdot\cdot\cdot)= D(\cdot\cdot\cdot) - \frac{\partial{(\cdot\cdot\cdot)}}{\partial y}dy.
\end{equation}
This definition is invariant under the set of transformations that keep unchanged the $4 + 1$ separation provided by (\ref{5D metric restricted metric}). For the case of more general metrics, $D^{(4)}$ can also be defined, but this requires the introduction of the appropriate projectors \cite{Ponce de Leon 1}. 

Physical quantities defined in $4D$ should be appropriately separated from their $5D$ counterparts. In particular, the $4D$ force (per unit mass) should be defined through $D^{(4)}u^{\mu}$ instead of $Du^{\mu}$, namely,
\begin{equation}
\label{correct definition of force}
f^{\mu}= \frac{D^{(4)}u^{\mu}}{ds}, \;\;\;\;\;\  f_{\mu}= \frac{D^{(4)}u_{\mu}}{ds}.
\end{equation}
Since $D^{(4)}g_{\mu\nu} = 0$, we have $f_{\sigma} = g_{\sigma\mu}f^{\mu}$, as desired.

Let us now find the contravariant components, $f^{\mu}$. Following (\ref{def of D in 4D}), $D^{(4)}u^{\mu} = Du^{\mu} - (\partial{u^{\mu}}/{\partial y})dy$. Thus, we need to evaluate $(\partial{u^{\mu}}/{\partial y})$. 
\begin{equation}
\label{calculation of partial derivative of u}
du^{\mu} = \frac{d(dx^{\mu})}{ds}-\frac{dx^{\mu}}{(ds)^2}d\left(\sqrt{g_{\alpha\beta}dx^{\alpha}dx^{\beta}}\right).
\end{equation}
Taking derivatives and rearranging terms
we get
\begin{equation}
\label{calculation of partial derivative of u, part 3}
\frac{\partial{u^{\mu}}}{\partial y}= - \frac{1}{2}u^{\mu}\frac{\partial{g_{\alpha\beta}}}{\partial y}u^{\alpha}u^{\beta}.
\end{equation}
For the covariant components $f_{\mu}$ we need $(\partial{u_{\mu}}/\partial y)$. This can be obtained from above and $u_{\mu} = g_{\mu\nu}u^{\nu}$, as
\begin{equation}
\label{partial derivative of u covariant}
\frac{\partial{u_{\mu}}}{\partial y} = \frac{\partial{g_{\mu\lambda}}}{\partial y}u^{\lambda} - \frac{1}{2}u_{\mu}\frac{\partial{g_{\alpha\beta}}}{\partial y}u^{\alpha}u^{\beta}.
\end{equation}
Collecting results, we finally have 
\begin{equation}
\label{general expression for my definition of force, contravariant}
\frac{D^{(4)}u^{(\sigma)}}{ds} = f^{\sigma}= 
 \left[u^{\sigma}u^{\lambda} - g^{\sigma\lambda}\right]u^{\rho} \frac{\partial{g_{\lambda\rho}}}{\partial y}\frac{dy}{ds}.
\end{equation}
Also,
\begin{equation}
\label{general expression for my definition of force, covariant}
\frac{D^{(4)}u_{\mu}}{ds} = f_{\mu} = \left[u_{\mu}u^{\rho} - {\delta}^{\rho}_{\mu}\right] u^{\lambda}\frac{\partial{g_{\rho\lambda}}}{\partial y}\frac{dy}{ds}.
\end{equation}
It follows that, with this new definition, the contravariant and covariant components of the force satisfy the usual requirements for four-vectors (\ref{self-consistency}). In particular, this force is orthogonal to the four-velocity of the particle.

Equations (\ref{general expression for my definition of force, contravariant})-(\ref{general expression for my definition of force, covariant}) show that the force from an extra non-compactified dimension does not necessarily contradict physics in $4D$. We propose these equations, instead of the abnormal force (\ref{f(lit) contr})-(\ref{relation between covariant and contravariant components of force in literature}), as the correct  expressions for the force from  a non-compactified extra dimension.

Finally, for completeness, we provide the equation for $(dy/ds)$. It is given by the fourth component of the geodesic equation as
\begin{equation}
\label{fourth component of the geod}
\frac{d^2y}{ds^2} = \frac{\epsilon}{2}\left[ 1 + \epsilon(\frac{dy}{ds})^2\right]\frac{\partial{g_{\mu\nu}}}{\partial y}u^{\mu}u^{\nu}.
\end{equation} 
We notice that $\epsilon$ does not appear explicitly in (\ref{general expression for my definition of force, contravariant})-(\ref{general expression for my definition of force, covariant}). However, the character of the extra dimension influences the $4D$ force via $(dy/ds)$, namely 
\begin{eqnarray}
\epsilon &=& -1,\,\,\,\, \frac{dy}{ds} = Tanh[\frac{1}{2}(w_{0} - w)]\nonumber \\
\epsilon &=& +1,\,\,\,\, \frac{dy}{ds} = tan[\frac{1}{2}(w - w_{0})],
\end{eqnarray}
where $w = \int{(\partial g_{\mu\nu}/\partial y)u^{\mu}u^{\nu}ds}$, and $w_{0}$ is a constant of integration.     

\section{Discussion and Conclusions}

The purpose of this work has been to show that the existence of an extra non-compactified dimension does not violate $4D$ laws of particle mechanics. 
With this aim, we have formulated a new definition  for the force from a non-compactified extra dimension, which is compatible with what we know in $4D$ physics (Eqs. (\ref{general expression for my definition of force, contravariant})-(\ref{general expression for my definition of force, covariant})). 

In order to get another perspective in the discussion, let us consider the so-called warp metrics. These are
\begin{equation}
\label{Warp metric}
d{\cal S}^2 =  \Omega(y) {\tilde{g}}_{\mu\nu}(x^{\rho}, y)dx^{\mu}dx^{\nu} + \epsilon dy^2,
\end{equation}
where the  conformal factor $\Omega$ is called {\it warp} factor, and 
 ${\tilde{g}}_{\mu\nu}(x^\rho, y)$ is interpreted as the {\it physical} metric on the embedded hypersurface of one lower dimension. These metrics are popular in ``brane" theory and space-time-matter theory \cite{J. Ponce de Leon 3}. In the case where $\tilde{g}_{\mu\nu}$ is {\em not } a function of $y$, the spacetime metric is essentially that of compactified Kaluza-Klein theory and we would {\em not}   expect any force from the extra dimension. However, a simple calculation from (\ref{f(lit) contr}) gives
\begin{equation}
\label{f lit for warp metric}
f_{(lit)}^{\mu} = - \frac{u^{\mu}}{2 \Omega}\frac{d\Omega}{dy}\frac{dy}{d{\tilde {s}}},
\end{equation} 
where $ds = \sqrt{\Omega} d\tilde{s}$ and now $u^\mu = dx^{\mu}/d{\tilde{s}}$. On the other hand, the calculation from (\ref{general expression for my definition of force, contravariant}) gives 
\begin{equation}
\label{fifth force for warp metrics}
f^{\mu} = 0,
\end{equation}
which is more acceptable from a physical point of view. Indeed, we would expect the force from the extra dimension should come from the dependence of the physical metric on $y$ and not from the conformal factor, as in (\ref{f lit for warp metric}). While this case is very simple, and more complicated metrics can be considered, it clearly illustrates our point. Namely that (\ref{general expression for my definition of force, contravariant}) is more consistent than (\ref{f(lit) contr}) not only with the usual physics in $4D$, but also with our physical intuition.

Predicting some effects of this new force will require some specific model. For astrophysical and cosmological observations/experiments, we can consider a line element with spherical symmetry 
\begin{equation}
\label{astr. metric}
d{\bf {\cal S}}^2 = e^{\nu(t, r, y)}(dt)^2 - e^{\lambda(t, r, y)}[(dr)^2 + r^2(d\Omega)^2] + \epsilon \Phi^2(t, r, y) (dy)^2,
\end{equation}
where $(d\Omega)^2 = (d\theta)^2 + sin^2\theta (d\phi)^2$, and the metric coefficients are some solution of the field equations. It is not difficult to see that the spatial part of $f^{\mu}$, in (\ref{general expression for my definition of force, contravariant}), is   collinear with the three-velocity of the particle. In short ${\bf f} = \alpha {\bf v}$, where $\alpha$ depends on $(\partial {\nu}/\partial y)$ and $(\partial {\lambda}/\partial y)$. Therefore the particle will move under the influence of two forces; the gravitational one (which roughly is proportional to $(\partial {\nu}/\partial r)$ and does not depend on the velocity) and the extra force which does depend on the velocity. 

One can imagine a scenario, of particles at high speed, where the extra force could be  comparable and even prevail over the gravitational one. The effects from this force could in principle be detected in ultra-relativistic particles in the vicinity of black holes and/or cosmological situations as the peculiar motions of galaxies \cite{Wesson Seahra}-\cite{Seahra Wesson}. 

 The implications of this force for astrophysics and cosmology is a topic worth of future investigation. This should give one the opportunity to test different models experimentally for their compatibility with observational data. 
\\
\\
\\
\\
{\bf Acknowledgment}\\

The author wishes to thank Sanjeev S. Seahra for interesting discussions.


\begin{thebibliography}{99}
\bibitem{Wesson book}{P.S. Wesson, {\em Space-Time-Matter} (World Scientific Publishing Co. Pte. Ltd. 1999).}
\bibitem{Youm}{D. Youm, hep-th/0004144 (2000), {\em Phys. Rev.} {\bf D62}, 084002(2000).}
\bibitem{Maartens}{R. Maartens, het-th/0004166 (2000), {\em Phys. Rev.} {\bf D62}, 084023(2000).}
\bibitem{Chamblin}{A. Chamblin, hep-th/0011128 (2000), {\em Class. Quant. Grav.} {\bf 18}, L17(2001).}
\bibitem{Deruelle and Katz}{N. Deruelle and J. Katz, gr-qc/0104007 (2001), {\em Phys. Rev.} {\bf D64}, 083515(2001).}
\bibitem{Wesson Mashhonn}{P.S. Wesson, B. Mashhoon, H. Liu, W.N. Sajko, {\em Phys. Lett.} {\bf B456}, 34(1999).}
\bibitem{RS2}{L. Randall and R. Sundrum, hep-th/9906064 (1999), {\em Phys. Rev. Lett.} {\bf 83}, 4690(1999).}
\bibitem{Youm 2}{Donam Youm, hep-th/0110013 (2001).}
\bibitem{Ponce de Leon 1}{J. Ponce de Leon, gr-qc/0104008 (2001).}
\bibitem{J. Ponce de Leon 3}{J. Ponce de Leon, gr-qc/0106020, {\em Mod. Phys. Lett.} {\bf A21}, 1405(2001).}
\bibitem{Wesson Seahra}{P.S. Wesson and Sanjeev S. Seahra, {\em Ap. J} {\bf 558}, L75(2001).}
\bibitem{Seahra Wesson}{Sanjeev S. Seahra and P.S. Wesson, {Gen. Relativ. and Gravit.} in press(2001).}
\end{thebibliography}
\end{document}